\documentclass[default]{sn-jnl}
\jyear{2021}%

\theoremstyle{thmstyleone}%
\theoremstyle{thmstyletwo}%

\theoremstyle{thmstylethree}%

\begin{document}

\title[]{A room-temperature moir{\'e} interlayer exciton laser}

\author*[1,3,4]{\fnm{Qiaoling} \sur{Lin}}\email{qiali@dtu.dk}
\equalcont{These authors contributed equally to this work.}
\author*[2]{\fnm{Hanlin} \sur{Fang}}\email{hanlin.fang@chalmers.se}
\equalcont{These authors contributed equally to this work.}
\author[5]{\fnm{Yuanda} \sur{Liu}}\email{liuyd@imre.a-star.edu.sg}
\author[6]{\fnm{Yi} \sur{Zhang}}\email{yi.1.zhang@aalto.fi}
\author[1,3,4]{\fnm{Moritz} \sur{Fischer}}\email{mofis@dtu.dk}
\author[7]{\fnm{Juntao} \sur{Li}}\email{lijt3@mail.sysu.edu.cn}
\author[8]{\fnm{Joakim} \sur{Hagel}}\email{joakim.hagel@chalmers.se}
\author[9]{\fnm{Samuel} \sur{Brem}}\email{samuel.brem@uni-marburg.de}
\author[9]{\fnm{Ermin} \sur{Malic}}\email{ermin.malic@uni-marburg.de}
\author[1,3,4]{\fnm{Nicolas} \sur{Stenger}}\email{niste@dtu.dk}
\author[6]{\fnm{Zhipei} \sur{Sun}}\email{zhipei.sun@aalto.fi}
\author[1,3,4]{\fnm{Martijn} \sur{Wubs}}\email{mwubs@dtu.dk}
\author*[1,3,4]{\fnm{Sanshui} \sur{Xiao}}\email{saxi@dtu.dk}

\affil[1]{\orgdiv{Department of Electrical and Photonics Engineering}, \orgname{Technical University of Denmark}, \orgaddress{\postcode{DK-2800}, \state{Kongens Lyngby}, \country{Denmark}}}
\affil[2]{\orgdiv{Department of Microtechnology and Nanoscience (MC2)}, \orgname{Chalmers University of Technology}, \orgaddress{\postcode{41296}, \state{Gothenburg}, \country{Sweden}}}
\affil[3]{\orgdiv{NanoPhoton - Center for Nanophotonics}, \orgname{Technical University of Denmark}, \orgaddress{\postcode{DK-2800}, \state{Kongens Lyngby}, \country{Denmark}}}
\affil[4]{\orgdiv{Centre for Nanostructured Graphene}, \orgname{Technical University of Denmark}, \orgaddress{\postcode{DK-2800}, \state{Kongens Lyngby}, \country{Denmark}}}
\affil[5]{\orgdiv{Institute of Materials Research and Engineering}, \orgname{Agency for Science Technology and Research (A*STAR)}, \orgaddress{\street{2 Fusionopolis Way}, \postcode{138634}, \country{Singapore}}}
\affil[6]{\orgdiv{Department of Electronics and Nanoengineering and QTF Centre of Excellence}, \orgname{Aalto University}, \orgaddress{\city{Espoo}, \postcode{02150}, \country{Finland}}}
\affil[7]{\orgdiv{State Key Laboratory of Optoelectronic Materials and Technologies, School of Physics}, \orgname{Sun Yat-Sen University}, \orgaddress{\city{Guangzhou}, \postcode{510275}, \country{China}}}
\affil[8]{\orgdiv{Department of Physics, \orgname{Chalmers University of Technology}, \orgaddress{\postcode{41296}, \state{Gothenburg}, \country{Sweden}}}}
\affil[9]{\orgdiv{Department of Physics, \orgname{Philipps-Universität Marburg}, \orgaddress{\postcode{35037}, \state{Marburg}, \country{Germany}}}}

\abstract{Moir{\'e} superlattices in van der Waals heterostructures offer highly tunable quantum systems with emergent electronic and excitonic properties such as superconductivity~\cite{cao2018unconventional}, topological edge states~\cite{tong2017topological}, and moir{\'e}-trapped excitons~\cite{tran2019evidence}. Theoretical calculations~\cite{yu2017moire,zhang2017interlayer} predicted the existence of the moir{\'e} potential at elevated temperatures; however, its impact on the optical properties of interlayer excitons (IXs) at room temperature is lacking, and the benefits of the moir{\'e} effects for lasing applications remain unexplored.
We report that the moir{\'e} potential in a molybdenum disulfide-tungsten diselenide (MoS$_2$/WSe$_2$) heterobilayer system can significantly enhance light emission, elongate the IX lifetime, and modulate the IX emission energy at room temperature. By integrating a moir{\'e} superlattice with a silicon topological nanocavity, we achieve ultra-low-threshold lasing at the technologically important telecommunication O-band thanks to the significant moir{\'e} modulation. Moreover, the high-quality topological nanocavities facilitate the highest spectral coherence of $<$ 0.1~nm linewidth among all reported two-dimensional material-based laser systems. Our findings not only open a new avenue for studying correlated states at elevated temperatures, but also enable novel architectures for integrated on-chip photonics and optoelectronics.}


\maketitle

\unnumbered 
\section{Introduction}\label{sec1}
Monolayer transition metal dichalcogenides (TMDs) have emerged as a new class of gain materials for lasing  applications~\cite{salehzadeh2015optically, ye2015monolayer, shang2017room, li2017room, ge2019laterally} due to their direct band gap, large exciton binding energy (on the order of 500~meV) and easy integration with various materials including the highly-developed silicon platform. The type-II band-aligned heterostructure with appropriately combined monolayer TMD gives rise to IXs, where the electrons and holes are spatially separated in different layers and bound by the Coulomb interaction. Compared to intralayer excitons in monolayer TMDs, IXs show much smaller oscillator strengths (around two to three orders of magnitude lower) and longer exciton lifetimes (ns instead of ps), thus enabling efficient IX accumulation and population inversion~\cite{paik2019interlayer,liu2019room}. Moreover, for such a heterostructure, it is by now widely recognized that moir{\'e} superlattices can form due to a small twist angle and lattice mismatch between two monolayers~\cite{huang2022excitons}. The appearance of moir{\'e} superlattices creates a lateral periodic potential for excitons, resulting in exciton localization (known as moir{\'e} excitons shown in Fig.~1a), which has been widely observed at cryogenic temperatures~\cite{tran2019evidence,jin2019observation, regan2022emerging}. From first-principle calculations~\cite{yu2017moire, tran2019evidence}, the depth of the moir{\'e} potential is predicted to be approximately 100-200~meV, much larger than the thermal energy of $\sim$27~meV at room temperature. A very recent experiment~\cite{sun2022enhanced} correlates the multipeak feature in IX emission spectra from free-standing WS$_2$/WSe$_2$ heterobilayers with moir{\'e} IXs, however still lacking strong evidence for the presence of moir{\'e} potential at room temperature. Furthermore, the additional in-plane quantum confinement in moir{\'e} superlattices could potentially enhance the quantum yield of light emission and lead to a large optical gain at low pumping levels, akin to quantum dots which showed lower lasing thresholds than quantum wells and bulk materials~\cite{ledentsov2000quantum}. However, the study of the moir{\'e} superlattice on laser performance has remained unexplored. 

Here we report the existence of room-temperature moir{\'e} excitons in MoS$_2$/WSe$_2$ heterobilayers by twist-angle-dependent photoluminescence (PL) measurements, reflection contrast spectroscopy, power-dependent measurements, and time-resolved PL dynamics. We find that the moir{\'e} potential gives rise to an ultra-wide emission tunability of IXs and strongly suppressed non-radiative recombinations. By integrating moir{\'e} superlattices with silicon topological photonic crystal nanocavities, to the best of our knowledge, we for the first time explore the potential of room-temperature moir{\'e} IX for high-performance laser applications. The combination of a high-quality-factor (Q-factor, $> 10^4$) cavities and moir{\'e} IX states leads to ultra-low-threshold lasing with emission wavelength extended to the optical fibre communication (OFC) O-band (1260-1360~nm), high side-mode suppression ratio (SMSR), and the highest spectral coherence among two-dimensional (2D) layered materials based lasers~\cite{wen2020excitonic} (Supplementary Table~S2). These findings encourage studying novel exciton physics in moir{\'e} superlattices at room temperature and open new avenues for using these artificial quantum materials in high-performance device applications.

\unnumbered 
\section{Results}\label{sec2}

\begin{figure}[h]
\centering
\includegraphics[scale=0.7]{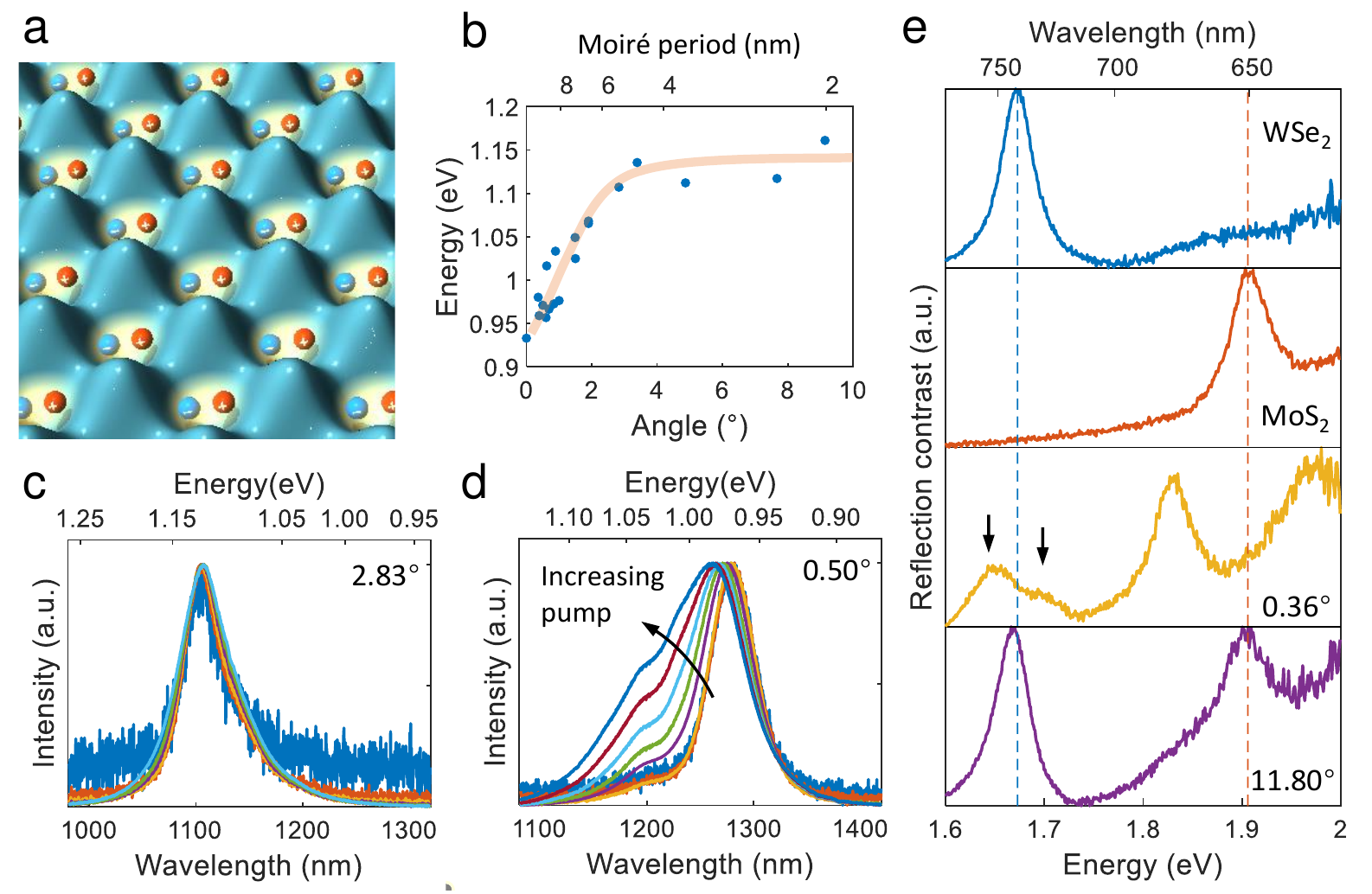}
\caption{Evidence of room-temperature moir{\'e} IXs in MoS$_2$/WSe$_2$ heterobilayers. 
(a) Schematic of IXs trapped in moir{\'e} potentials. 
(b) IX peak emission energy versus twist angle. The light red line is a guide to the eye. The twist angles are the relative rotation angles between the monolayers.
Power-dependent PL spectra of heterobilayers with twist angles of 2.83$^\circ$ (c) and 0.50$^\circ$ (d) up to the same maximum power level. For the heterobilayer with a small twist angle in (d), the evolution with power exhibits a strong blue shift and the emergence of high-energy states, indicated by the black arrow. 
(e) Reflection contrast spectra of heterobilayers with twist angles of 0.36$^\circ$ (yellow) and 11.80$^\circ$ (purple), together with spectra of reference monolayers WSe$_2$ (blue) and MoS$_2$ (red). As a reference, the vertical blue (red) dashed line presents A exciton peaks for monolayer WSe$_2$ (MoS$_2$). The multi-peak appearance (two black arrows) indicates the existence of the moir{\'e} effect when the twist angle is 0.36$^\circ$ (yellow).
}
\label{fig.1}
\end{figure}

In this work, we use MoS$_2$/WSe$_2$ heterobilayers (HB) for their bright K-K transition of IXs with emission energy below 1.1~eV~\cite{karni2019infrared}, which is compatible with well-developed silicon photonics. Monolayer MoS$_2$ and WSe$_2$ are exfoliated mechanically and stacked with the dry transfer technique (details in Methods)~\cite{zomer2014fast} to obtain a clean interface and high-quality optical properties. We prepare more than 25 MoS$_2$/WSe$_2$ heterobilayer samples with various twist angles and all measurements are carried out at room temperature. Fig.~1b shows the peak energy of IX emission as a function of the twist angle $\theta$. The exciton energy changes significantly with the twist angle ($>$200~meV), in particular at angles below 2$^{\circ}$. This pronounced energy shift is similar to that observed in MoSe$_2$/WSe$_2$ heterobilayers at cryogenic temperatures, where the large energy shift is ascribed to the moir{\'e} potential on top of the twist angle-dependent momentum mismatch~\cite{barre2022optical}. 

We further perform power-dependent measurements for two representative samples with different twist angles up to the same pumping level. For a heterobilayer of $\theta=2.83^{\circ}$ (Fig.~1c), the IX has an emission energy of $\sim$1.12~eV ($\sim$1107~nm) and this energy remains constant with increasing power. In stark contrast, a remarkable blue-shift and the emergence of high-energy states are observed when $\theta=0.50^{\circ}$ (Fig.~1d). We attribute these to the existence of the moir{\'e} potential that acts as an exciton reservoir for different exciton states and enhances the exciton-exciton interaction. 
As the potential confines excitons laterally resulting in quantized energy level~\cite{tran2019evidence}, IXs obtain a larger chance to fill higher-energy moir{\'e} excitonic states when the pumping power increases. The filling to higher-energy states leads to the occurrence of the blue shift for the emission envelope. We also report the observation of spectrally resolved moir{\'e} states in hexagonal boron nitride (hBN) encapsulated heterobilayers, which is shown in Supplementary fig.~S1 and will be discussed later.

The moir{\'e} potential not only traps IXs but also affects intralayer excitons~\cite{jin2019observation}. Here we present another piece of evidence for the existence of moir{\'e} potentials at room temperature by characterizing the absorption properties of intralayer excitons using reflection contrast spectroscopy. Fig.~1e presents the reflection contrast spectra, where the reference signals of monolayer WSe$_2$ and MoS$_2$ are shown in blue and red, respectively. We focus on the WSe$_2$ A exciton in the range between 1.6 and 1.75 eV. The heterobilayer with the twist angle of $\sim$0.36$^{\circ}$ (yellow) shows an emergent doublet (highlighted by two arrows in Fig.~1e) with comparable oscillator strengths near the WSe$_2$ exciton (1.675~eV). By contrast, the heterobilayer with the large twist angle ($\sim$11.80$^{\circ}$) shows a single exciton resonance. The appearance of multiple emergent peaks around the original WSe$_2$ exciton serves as strong evidence for the presence of moir{\'e} excitons at room temperature. It has already been observed at low temperatures~\cite{jin2019observation,stansbury2021visualizing,karni2022structure} and is well explained by the theoretical model where the moir{\'e} potential generates multiple flat moir{\'e} minibands~\cite{jin2019observation}.

\begin{figure}[h]
\centering
\includegraphics[scale=0.75]{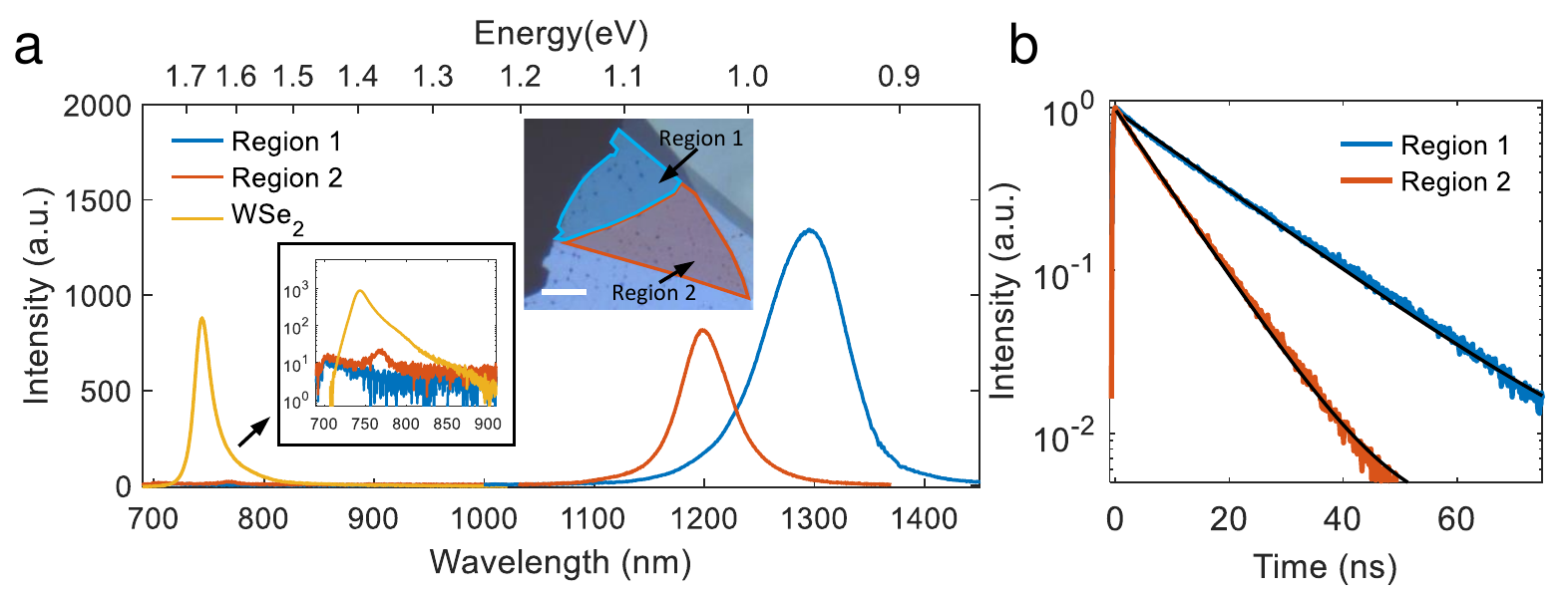}
\caption{Moir{\'e} modulated IX emission. 
 (a) PL spectra from Region~1 (blue), Region~2 (red) and a reference monolayer WSe$_2$ (yellow). The PL spectra of the intralayer exciton on a logarithmic scale (black frame inset) show the higher interlayer coupling in Region~1. Inset: the optical image of the heterobilayer consisting of two artificially colored regions. Scale bar: 10~$\mu $m.
 (b) Time-resolved PL dynamics from two regions for IXs emission longer than 900~nm. The black curves are exponential fits to the data, giving an IX lifetime of 17.6~ns (8.5~ns) for Region~1 (Region~2).
}
\label{fig.2}
\end{figure}

Recent studies clearly show that the atomic registry of the moir{\'e} superlattice dictates the behaviour of interlayer electronic coupling~\cite{zhang2017interlayer}, which can be used as a tuning knob for novel 2D electronic systems. Here we examine the influence of the moir{\'e} potential on the IX lifetime that plays a crucial role in population inversion. The twist angle has a substantial impact on the IX lifetime, as it generates not only a relative shift in momentum space but also modifies the moir{\'e} supperlattices~\cite{choi2021twist}. However, it is challenging to differentiate the contributions to radiative decay from the momentum mismatch (without moir{\'e} potential) and the moir{\'e} effect. To largely exclude the effect of the momentum mismatch, we choose a representative near-zero-twist-angle heterobilayer shown as the inset in Fig.~2a, where the optical image shows two clearly distinguishable separated regions. These two regions (Region~1 and Region~2) can be identified by their different optical contrast, implying the presence of different interlayer coupling while simultaneously having the same twist angle. The PL spectra shown in Fig.~2a cover the WSe$_2$ exciton emission ($\sim$750~nm) and IX emission ($\sim$1000-1400~nm). Both regions feature strong interlayer coupling supported by the approximately two orders of magnitude emission quenching of the WSe$_2$ A excitons. On the logarithmic scale (see the inset with black frame), the different intralayer exciton intensities from the two regions further indicate that Region~1 has a better coupling than Region~2. Compared to Region~2, the IX emission from Region~1 exhibits a brighter emission intensity and lower emission energy (see the right part of Fig.~2a). These results can be explained by: (1) Stronger interlayer coupling, i.e., the increased overlap of the wavefunctions of the spatially separated electrons and holes, results in a larger probability of IX radiative recombination. (2) The interlayer coupling determines the depth of the moir{\'e} potential~\cite{tran2019evidence}, and stronger interlayer coupling leads to a deeper moir{\'e} potential which thus can trap IXs in lower energy states. The increased radiative recombination due to the strong interlayer coupling usually leads to fast decay. 
However, our time-resolved PL measurements (Fig.~2b) show that the IXs in Region~1 feature a longer lifetime ($\sim$17.6~ns) than Region~2 ($\sim$8.5~ns). This counterintuitive result can be explained by the strong suppression of non-radiative recombination at elevated temperatures due to the spatial confinement of IXs with moir{\'e} potential~\cite{sirigu2000excitonic} that impedes the exciton diffusion to non-radiative centers~\cite{choi2020moire, li2021interlayer}. 
We conclude that the moir{\'e} potential not only enhances the light emission of IX but also prolongs its lifetime, which is favourable for lasing applications.\\

\begin{figure}[h]
\centering
\includegraphics[scale=0.7]{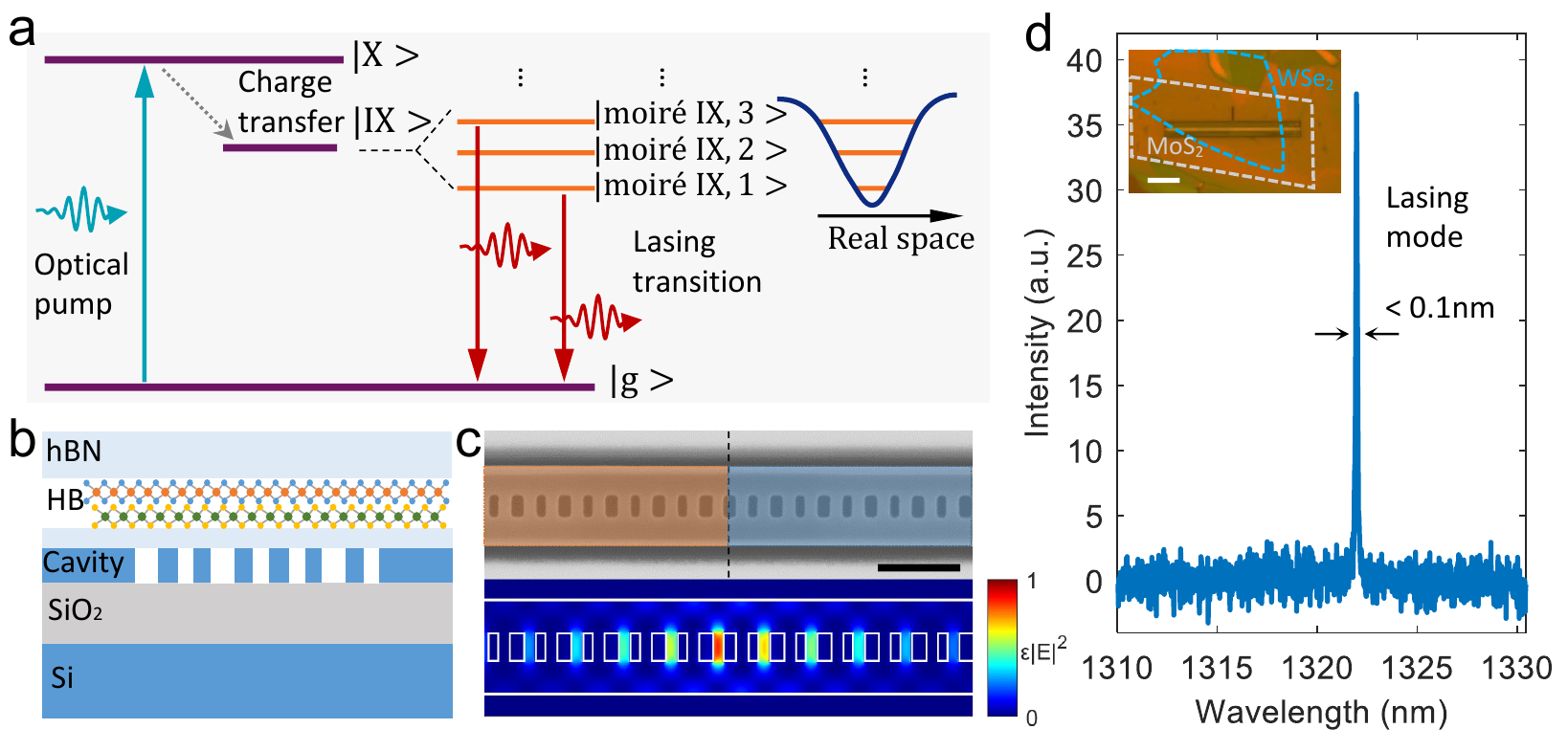}
\caption{Demonstration of moir{\'e} IX lasers. 
(a) Energy level diagram of moir{\'e} IX states for lasing emission.
(b) Schematic of the hBN-encapsulated  MoS$_2$/WSe$_2$ heterobilayer integrated with a SiO$_2$-supported silicon topological nanobeam cavity. 
(c) SEM image (the upper panel) of the cavity. The light red and blue areas illustrate two photonic crystals with different Zak phases, leading to the appearance of a cavity mode at their interface. The simulated mode profile is illustrated in the lower panel. Scale bar, 500~nm. 
(d) The emission spectrum of Device~1 at an excitation power intensity of $\sim$4.39~kW/cm$^2$. Inset: Optical microscope image of the device. The blue and grey dashed boxes outline the monolayer WSe$_2$ and MoS$_2$ areas. The top and bottom hBN layers are not shown. Scale bar, 5~$\mu$m.
}
\label{fig.3}
\end{figure}

To explore the potential of moir{\'e} IX for laser applications, a rotationally aligned MoS$_2$/WSe$_2$ moir{\'e} superlattice is integrated with a silicon photonic crystal cavity. An equivalent three-level system considering the multiple moir{\'e} states and a sketch of the device are shown in Figs. 3a and 3b, respectively. Intralayer exciton states $\mid$X$>$ are first generated after optical pumping and follow an ultrafast charge transfer process to form IX states $\mid$IX$>$. The moir{\'e} potential leads to the formation of quantized IX energy levels, marked as $\mid$moir{\'e} IX,1$>$,$\mid$moir{\'e} IX,2$>$,$\mid$moir{\'e} IX,3$>$, etc. The heterobilayer is encapsulated within two hBN flakes to prevent optical performance degradation in ambient environments~\cite{fang2019laser} and to suppress inhomogeneous linewidth broadening~\cite{tran2019evidence}. 
Thanks to this hBN encapsulation, one can observe a clear multipeak feature in PL spectra (see Supplementary fig.~S1), reinforcing the presence of moir{\'e} IX states.
Photonic topological cavities are chosen here due to their demonstrated robust single-mode operation~\cite{bandres2018topological, ota2018topological}, preventing mode competition thus simplifying the study of moir{\'e} IX-cavity interaction. The inset in Fig.~3d presents the optical image of Device~1 fabricated by standard e-beam lithography and polymer-based transfer technique~\cite{zomer2014fast} (details in Methods). 
The top panel in Fig.~3c shows the scanning electron microscopy (SEM) image of our fabricated topological cavity, and the simulated electric energy distribution shown in Fig.~3b (the bottom panel) illustrates the single cavity mode localized at the interface between two photonic crystals with distinct Zak phases~\cite{ota2018topological}. We highlight that the Q-factor ($> 10^4$) of our cavity is one order of magnitude larger compared to the previous heterobilayer-based IX lasers~\cite{paik2019interlayer,liu2019room}, offering a low-loss reservoir for the photons thus enhancing coherent emission.

\begin{figure}[h]
\centering
\includegraphics[scale=0.7]{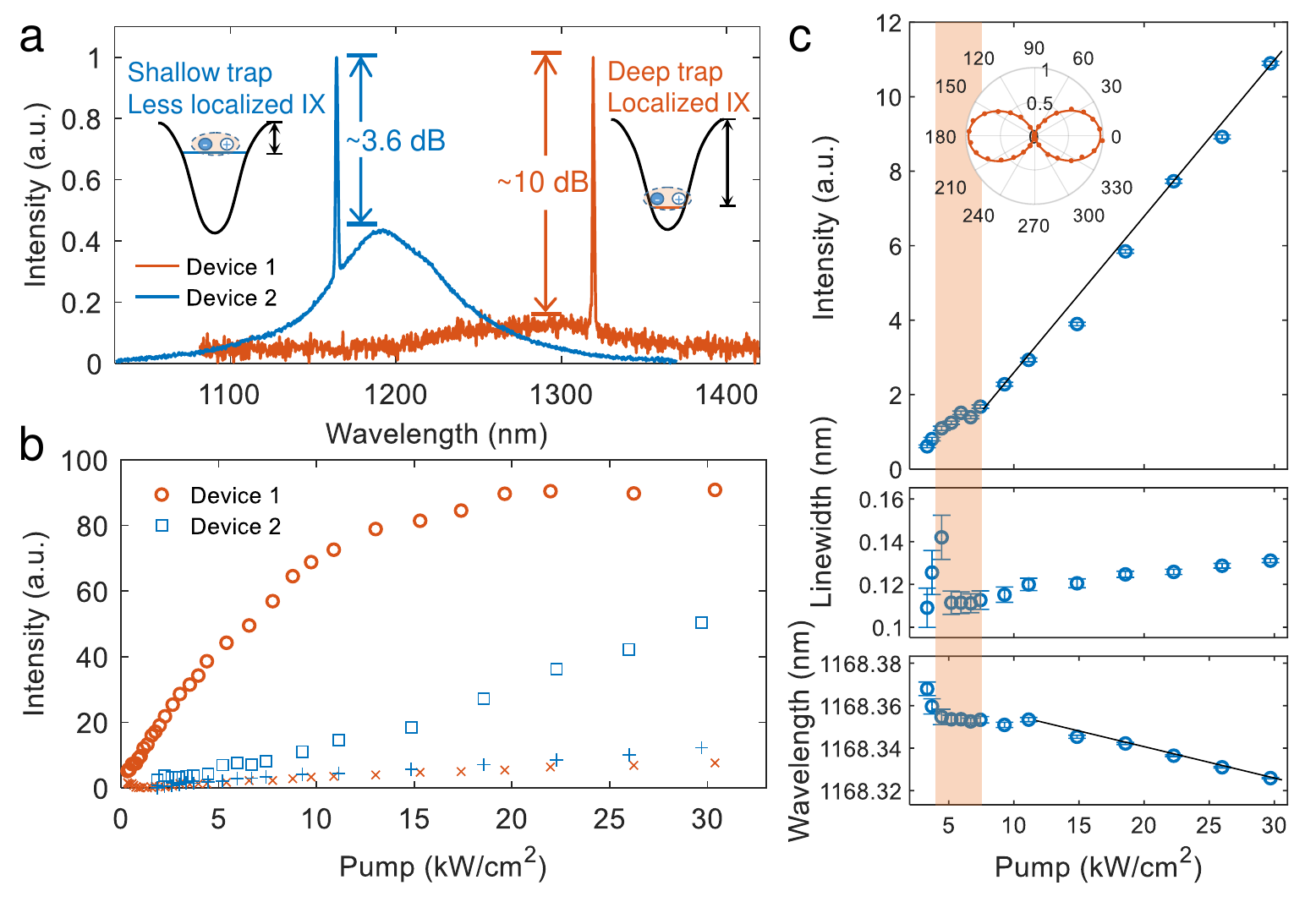}
\caption{Signatures of high-performance lasing.
(a) Two types of lasing operation: Device~1 (with localized moir{\'e} IX) and  Device~2 (with less localized moir{\'e} IX) at an excitation power intensity of $\sim$4.04~kW/cm$^2$ and $\sim$18.12~kW/cm$^2$ respectively. The black lines indicate the effective moir{\'e} potential depth. 
(b) Output behaviour of two devices. The hollow circle/square (cross) represents the lasing emission (spontaneous emission).
(c) Integrated intensity, linewidth, and wavelength of the output of Device~2 as a function of pump intensity. The experimental data are fitted by a Lorentzian function, and the error bars correspond to the 95\%  confidence interval of the Lorentzian fit. The threshold region is highlighted by the light red shaded area and the inset shows the linear polarization of the emitted mode characterized by polarization-resolved measurements.
}
\label{fig.4}
\end{figure}

As displayed in Fig.~1b, the presence of moir{\'e} potential extends the light emission of IX to the OFC O-band, which is essential for high-speed data transmission and other applications. By coupling the deeply trapped IX to the cavity mode, we observe a pronounced sharp emission line at $\sim$1321~nm (Fig.~3d) under optical pumping, denoted as Device~1. It should be noted that the measured emission linewidth of $\sim$0.1~nm is limited by the spectral limit of our spectrometer with high-resolution grating (1200~grooves/mm), showing the highest spectral purity of our device among the reported 2D material lasers (see the comparison in Supplementary Table~S2). This narrow linewidth corresponds to a coherence time of $\sim$45~ps (see the formula in Methods), one order of magnitude longer than the previously reported IX laser~(1.7~ps)~\cite{liu2019room}, thus giving strong evidence for highly coherent light emission. To obtain more details about the optical coupling between moir{\'e} IX and cavity resonance, a low-resolution grating (150~grooves/mm) is used to acquire a broad emission spectrum that includes the moir{\'e} IX emission and lasing mode (see the red curve in Fig.~4a). It is obvious that a single lasing mode dominates the spectrum, which confirms the non-trivial property of the topological cavity. In addition, the lasing wavelength locates at the low-energy part of the spontaneous emission peak of moir{\'e} IX with an SMSR up to $\sim$10~dB that is significantly larger than the typical value ($\sim$3-4~dB, shown in Supplementary Table~S2) of the reported 2D material lasers~\cite{liu2019room}. Interestingly, we find that the output intensity of Device~1 is linearly increased with pump intensity (Fig.~4b) and the lasing mode keeps dominating the emission even at low pumping levels. We believe that a large excitonic gain at OFC O-band is obtained due to the shift of the absorption peak of moir{\'e} IXs, which gives rise to a low lasing threshold beyond the measurement capability of our instruments.\\
For comparison, we design another device (Device~2) where the cavity resonance is mainly coupled to high-energy moir{\'e} IXs (Supplementary fig.~S1). In this case, the effective potential for IXs is shallower (left inset of Fig.~4a) which is analogous to the previously reported delocalized IX laser\cite{liu2019room}, and the lasing mode starts to dominate the emission spectrum at high pumping levels. With the increased excitation power, the evolution of the emission spectrum of Device~2 is investigated (Fig.~4c). The L-L curve (i.e., output intensity as a function of input power) presents a super-linear behaviour, which is a typical signature of lasing behaviour. In addition, a clear narrowing of the emission linewidth is observed near the threshold ($\sim$5~kW/cm$^2$), suggesting the occurrence of a phase transition from thermal to coherent emission. It should be noted that the observed threshold is reproducible (see Supplementary fig.~S3). Below the threshold, we obtain a Q-factor of the loaded cavity around $10^4$, indicating a low-loss laser system that allows us to reach the lasing condition with a relatively small optical gain. This can be attributed to the high Q factor of the passive cavity and the negligible absorption loss of IXs. When pumping at higher levels ($>$ 10~kW/cm$^2$), the lasing mode is blueshifted and accompanied by linewidth broadening (Fig.~4c), which can be explained by free-carrier absorption in the silicon cavity~\cite{zhao2014blue}. Our device barely suffers from pumping-induced heating as observed in previous works~\cite{li2017room, fang20181305}, thanks to the supported cavity design and hBN encapsulation enabling the efficient thermal energy dissipation~\cite{cai2019high}. Furthermore, the device exhibits excellent linear polarization with a degree of polarization (defined in Methods) of $\sim$0.94 of the emission line (inset of Fig.~4a), showing the efficient coupling between moir{\'e} IX and cavity mode. These observations indicate the presence of moir{\'e} excitonic gain for lasing. Furthermore, Fig.~4b exhibits that the output intensity of Device~1 grows faster (i.e., larger slope efficiency) than Device~2, indicating the existence of a high-gain medium that amplifies the light emission more efficiently. This can be further confirmed by the higher output intensity and lower saturation pumping power of Device~1. These findings prove the lower-threshold feature of Device~1 and suggest a relatively larger optical gain of the low-energy moir{\'e} IX states. The observation of lasing emission at 1321~nm and 1168~nm suggests a broad gain spectrum of the moir{\'e} excitons. This can possibly be ascribed to enhanced IX-IX interactions in our system arising from the additional quantum confinement by the moir{\'e} potential that acts as a reservoir for the accumulation of IXs.

\unnumbered 
\section{Discussion}\label{sec3}
In summary, we demonstrate the impact of moir{\'e} potentials on the light emission of IX at room temperature and their great potential for high-performance nanolasers. The combination of moir{\'e} IXs and high-Q silicon topological nanocavities results in ultra-coherent light emission with a low threshold. We emphasize that due to low signal it is challenging to perform photon correlation measurements, which can provide further evidence for laser operation~\cite{kreinberg2017emission}.
We note that a complete description of the moiré IX lasing behaviour requires a sophisticated model that fully accounts for the gain mechanisms in the moir{\'e} superlattices, which is beyond the scope of this study. Although the direct measurement of IXs absorption in TMDs at low temperatures has recently been achieved with electromodulation spectroscopy~\cite{barre2022optical}, further development of absorption measurement techniques at room temperature is strongly desired for determining the gain mechanisms in moir{\'e} superlattices. The discovery of moir{\'e} exciton gain at room temperature opens up the possibility of high-performance optoelectronic devices.

\section{Methods}\label{sec4}
\subsection*{Sample fabrication}\label{subsec1}
TMD monolayers were prepared by mechanical exfoliation from flux-grown bulk crystals (2D semiconductors) using polydimethylsiloxane (PDMS). The layer thickness was measured by optical contrast and PL emission energy. The heterobilayers were aligned by the straight edges of the TMD monolayers and stacked with the dry transfer technique~\cite{zomer2014fast}. The twist angles are determined based on the straight edges of the stacked heterobilayers as reported in previous work~\cite{baek2020highly} and are consistent with our second harmonic generation (SHG) measurements (Supplementary fig.~S2).
\subsection*{Device fabrication}\label{subsec2}
The topological photonic nanobeam cavities were fabricated on a commercial 220~nm-SOI wafer with a 2~$\mu$m sacrificial silicon dioxide layer (SOITEC). A 180~nm layer of e-beam resist (CSAR AR-P6200 ) was prepared, followed by an electron beam lithography process (JBX-9500FSZ, JEOL). After development, the cavity was created via an inductively coupled plasma dry-etching process. The residual resist was finally dissolved in 1165 solvent. 
The hBN thin films were exfoliated onto a SiO$_2$/Si substrate for encapsulation. The bottom hBN layer was kept thinner than 10~nm for good coupling between moir{\'e} IXs and the cavity mode. The hBN-encapsulated MoS$_2$/WSe$_2$ heterostructures were assembled and transferred to the nanocavities using the polymer-based dry-transfer method~\cite{zomer2014fast}. 
\subsection*{Optical measurements}\label{subsec3}
All measurements were performed at room temperature. For the PL measurements, a 637~nm continuous-wave laser diode was used to pump the devices. A 50x objective lens with a numerical aperture (NA) of 0.65 (LCPLN50XIR, Olympus) was used for both excitation and collection. The laser light was blocked through longpass spectral filters, and the PL signals were sent to a Czerny-Turner monochromator (SR500i, Andor) with a cooled InGaAs 1D-array camera (DU491A-1,7, Andor). A 150~grooves/mm grating is used for the wide-range light emission from moir{\'e} IXs (shown in Figs.~1c, 1d, 2a and 4a). For a better spectral resolution, the lasing mode was detected via a 600-grooves/mm grating (with a resolution of 0.22~nm) or a 1200~grooves/mm grating (0.1~nm).
The reflection contrast spectra were measured by using a white light source (SuperK Compact).
The time-resolved PL measurements were performed using a time-correlated single-photon counting (TCSPC) technique with a time tagger. We excited the device with a 640~nm pulsed laser (LDH-IB-640-B, PicoQuant) with a pulse width of $<90$~ps and a repetition rate of 10~MHz. The signals with wavelengths longer than 900 nm were sent to a single-photon detector (id220-FR, iDQ).
Polarization-dependent SHG measurements were used to determine the twist angle of heterobilayers and were carried out with an excitation wavelength of 960 nm (repetition rate 2 kHz) from an amplified Ti:sapphire femtosecond laser system (Spectra-Physics Solstice Ace). The polarization orientation of the excitation beam was tailored by rotating a half-wave plate (HWP). The laser light after the HWP was focused onto the sample by a 40x objective lens (NA=0.75, Nikon). The transmitted SHG signal was collected by another 40x objective lens (NA=0.5, Nikon), and passed through a linear polarizer. A 700-nm short-pass filter was placed after the polarizer to cut off the excitation beam. The final signal was detected by a photomultiplier tube (PMT) (Hamamatsu). Through the SHG measurements, the measured twist angle was correlated with IX energy (Supplementary fig.~S2).
\subsection*{Numerical simulation}\label{subsec4}
The topological photonic nanobeam cavity that we adopted was first proposed in Ref.~\cite{ota2018topological}. We retained a SiO$_2$ layer to support silicon nanobeams as our previous design ~\cite{fang2019laser} that was mechanically robust for the 2D material transfer process and was preferable for efficient heat dissipation. We optimized the cavity mode and parameters to reach a desired resonant wavelength using the finite-difference time-domain method (Lumerical, Ansys). The detailed structure parameters of the cavities are shown in Supplementary Table 1. 

\subsection*{Coherence time}\label{subsec5}
The coherence time was estimated by the formula~\cite{deng2002condensation}: $\tau_{\rm c} =\sqrt{8 \rm ln2} \lambda ^{2} /\left ( c\Delta \lambda  \right )$, where $\lambda $ is wavelength, $\Delta \lambda$ is spectral linewidth.

\subsection*{Degree of polarization}\label{subsec6}
The degree of polarization is defined as $(I_{\rm{max}}-I_{\rm{min}})/(I_{\rm{max}}+I_{\rm{min}})$, where $I_{\rm{max}}$ ($I_{\rm{min}}$) is the maximum (minimum) output intensity.
% \bmhead{Supplementary information}

\bmhead{Acknowledgments}
We thank Witlef Wieczorek for helpful discussions. 
This work was partially funded by the Danish National Research Foundation through the Center for Nanostructured Graphene (project no. DNRF103) and through NanoPhoton - Center for Nanophotonics (project no. DNRF147). N.S. acknowledges the support from the VILLUM FONDEN (project no. 00028233). Q.L. and S.X. acknowledge the support from the Independent Research Fund Denmark (project no. 9041-00333B and 2032-00351B), Direktør Ib Henriksens Fond, and Brødrene Hartmanns Fond. N.S. and M.W. acknowledge the support from the Independent Research Fund Denmark, Natural Sciences (project no. 0135-00403B). H.F. acknowledges support by the Olle Engkvists stiftelse, the Carl Tryggers stiftelse, and Chalmers Excellence Initiative Nano. J. L. and H. F. acknowledge the support from the National Natural Science Foundation of China (project no. 11974436). Y.L. acknowledges the support from A*STAR Career Development Fund - Seed Projects (C222812008). Y. Z. and Z. S. acknowledge the support from Horizon Europe (HORIZON) Project: ChirLog (101067269), the Academy of Finland (grants 314810, 333982, 336144, 336818, 352780 and 353364), Academy of Finland Flagship Programme (320167, PREIN), the EU H2020-MSCA-RISE-872049 (IPN-Bio), and ERC advanced grant (834742).

\bmhead{Author contributions}
H. F. and S. X. conceived and supervised the project. Q. L. and H. F. designed the device. Q. L., J. L., and Y. L. fabricated the devices. Q. L. and H. F. performed optical characterizations. Y. Z. performed the polarization-resolved SHG measurements. M.F. helped with lifetime measurements. Q. L. and H. F. carried out simulations about topological cavities. J. H. and E. M. assisted with the experiments. Q. L., H. F., Z. S., M. W., N. S., and S. X. analyzed the data. All authors contributed to the discussion and writing of the manuscript.

% \bibliographystyle{naturemag}
% \bibliography{references}

\end{document}